\documentclass[conference]{IEEEtran}
\IEEEoverridecommandlockouts
\usepackage{cite}
\usepackage{amsmath,amssymb,amsfonts}
\usepackage{algorithmic}
\usepackage{graphicx}
\usepackage{textcomp}
\usepackage{xcolor}
\usepackage{balance} 
\usepackage{tabularx, booktabs}
\usepackage{enumitem}
\usepackage{url}
\def\BibTeX{{\rm B\kern-.05em{\sc i\kern-.025em b}\kern-.08em
    T\kern-.1667em\lower.7ex\hbox{E}\kern-.125emX}}
\begin{document}

\title{Security Analysis and Threat Modeling of Research Management Applications [Extended Version]\\{\footnotesize \textsuperscript{*}Note: This is an extended version of a paper published in IEEE SoutheastCon 2025. © 2025 IEEE}\thanks{This is an extended version of a paper published in IEEE SoutheastCon 2025. © 2025 IEEE.}}

\author{\IEEEauthorblockN{Boniface M. Sindala}
\IEEEauthorblockA{\textit{Department of Computer Science} \\
\textit{University of Alabama at Birmingham}\\
Birmingham, Alabama, USA \\
bsindala@uab.edu}
\and
\IEEEauthorblockN{Ragib Hasan}
\IEEEauthorblockA{\textit{Department of Computer Science} \\
\textit{University of Alabama at Birmingham}\\
Birmingham, Alabama, USA \\
ragib@uab.edu}
}

\maketitle

\begin{abstract}
Research management applications (RMA) are widely used in clinical research environments to collect, transmit, analyze, and store sensitive data. This data is so valuable making RMAs susceptible to security threats. This analysis, analyzes RMAs' security, focusing on Research Electronic Data Capture (REDCap) as an example. We explore the strengths and vulnerabilities within RMAs by evaluating the architecture, data flow, and security features. We identify and assess potential risks using the MITRE ATT\&CK framework and STRIDE model. We assess REDCap's defenses against common attack vectors focusing on security to provide confidentiality, integrity, availability, non-repudiation, and authentication. We conclude by proposing recommendations for enhancing the security of RMAs, ensuring that critical research data remains protected without compromising usability. This research aims to contribute towards a more secure framework for managing sensitive information in research-intensive environments.
\end{abstract}

\begin{IEEEkeywords}
Threat Model, RMAs, REDCap, Security 
\end{IEEEkeywords}

\section{Introduction}
\label{sec:intro}
Research management applications (RMAs) are increasingly prevalent, transforming how research institutions manage data, foster collaboration, and ensure compliance~\cite{harris2019redcap}. They are essential tools in research enabling researchers to organize projects, collect data, and collaborate across institutions efficiently. They provide robust data collection, management, and analysis solutions. Developed by Vanderbuilt University, Research Electronic Data Capture (REDCap) is a widely used web-based RMA that facilitates capturing of research data and has become a cornerstone in clinical and translational research. Clinical trials, retrospective studies, and cohort studies, are some of the research types facilitated by REDCap~\cite{garcia2021research}.

However, with their widespread and nature of data they manage, security is paramount to ensure RMAs are not compromised. RMAs are subject to cyber-security vulnerabilities which can expose their critical functions. These vulnerabilities can be used to access sensitive data, that is collected, transmitted, and managed by RMAs. RMAs are prime targets because clinical data is more valuable than even credit card data~\cite{ghubaish2020recent}.  RMAs can be attacked in various ways which need protection such as devices used for access, the network, and data storage systems and thus exposing sensitive data~\cite{zisad2024towards}. This is underscored by the current threat landscape, which has seen unprecedented data compromise. For instance, the \textbf{Change Healthcare} breach in 2024 exposed the Protected Health Information (PHI) of an estimated of over 190 million individuals, making it the largest healthcare data breach ever reported~\cite{sharp2025health}. This scale of attack, stemming from a compromised Citrix remote access portal that lacked Multi-Factor Authentication (MFA), reinforces the urgent need for a robust security analysis that considers both systemic vulnerabilities and advanced attack capabilities. Injection of malware into the system would allow attackers to manipulate data to their liking or restrict access for legitimate users compromising patient safety and the privacy of research data. Data safety and privacy is paramount. RMAs security is a very critical component and should be implemented thoroughly considering all possible threats, focusing on data integrity, confidentiality, and compliance with regulatory bodies. 

To assess and comprehend the security threats impacting RMAs and develop mitigation strategies, we employed a systematic approach called threat modeling which helps in identifying, quantifying, and mitigating security threats~\cite{mahak2021threat}, helping us analyze RMAs from a security perspective. They are a variety of vulnerability assessment methods that can be used. In our research, we explore and investigate potential risks and vulnerabilities that RMAs may face within research institutions using the MITRE ATT\&CK framework~\cite{strom2018mitre} and STRIDE methodology~\cite{khan2017stride}. These models help us discover and define potential vulnerabilities and develop effective mitigation strategies to keep sensitive data secure and reduce risk. We use Data Flow Diagrams (DFDs) to identify possible vulnerable links during collection, transmission, analyzing, and storage of data, and develop mitigation strategies to better protect data within those links. We focus on REDCap as an example, but this research is intended to provide strategies to protect RMAs in general and to professionals and health care facilities for identifying and mitigating vulnerabilities and updating cyber-security measures. This helps to ensure that the security and trustworthiness of RMAs is maintained, especially considering that we are in an ever evolving cyber-security landscape.

\noindent\textbf{Contributions:} This research paper's contributions are:
\setlength{\leftmargini}{15pt}
\begin{enumerate}
    \item We developed a comprehensive threat model for research management applications (RMAs).
    \item We identified potential threats and vulnerabilities using the MITRE ATT\&CK framework and STRIDE methodology.
    \item We proposed mitigation strategies through analyzing threats and vulnerabilities to ensure that RMAs remain secure and sensitive data uncompromised.
\end{enumerate}

\textbf{Organization:} After the abstract and introduction, Section II presents RMA Background and threat model. Section III defines RMA assets. Section IV probes attack points of entry. Section V probes the attack model, attackers, and their motivation. Section VI explores RMA threats and vulnerabilities. Section VII explores mitigation strategies. Section VIII discuses related works and Section IX is the conclusion.

\section*{New Contributions in This Extended Version}
In this extended version\footnote{
This paper is an extended version of our previously published work at IEEE SoutheastCon 2025, with additional sections on the Dolev-Yao threat model, current trends in RMA security, and updated case studies.
}, what have added:
\begin{itemize}
  \item Updated case studies and vulnerabilities (e.g., Change Healthcare breach in 2024).
  \item Discussion of current trends including ethical requirements, data minimization, collaboration and traceability, and attack model.
  \item Integration of the Dolev-Yao threat model.
  \item Current trends in mitigation such as Zero Trust Architecture (ZTA)
\end{itemize}

\section{Background}
\label{sec:background}
\subsection{Research Management Applications}
\begin{figure}[!t]
    \centerline{\includegraphics[width=2.8in,alt={RMA Workflow}]{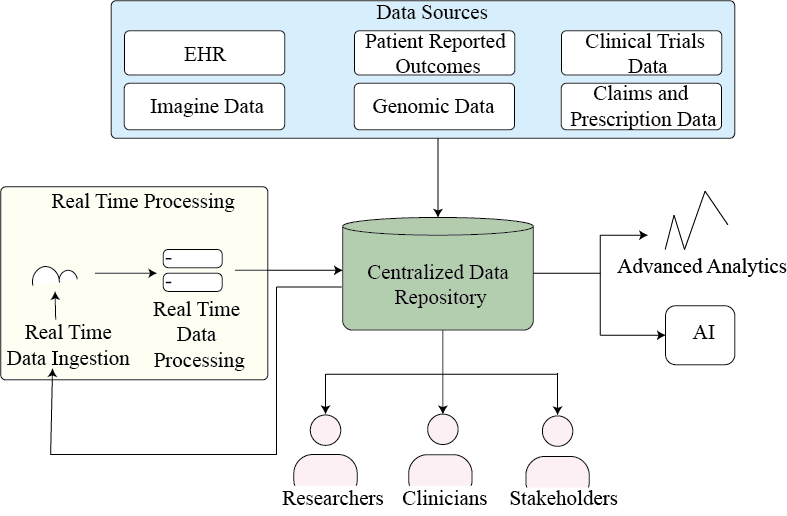}}
    \caption{Research Management Application workflow}
    \label{fig:rma}
\end{figure}

RMAs streamline research workflows by organizing, collecting, analyzing, and storing data. They facilitate collaboration within research institutions and manage the complexities of research protocols and participant management, ensuring compliance with industry regulatory bodies like HIPAA and General Data Protection Regulation (GDPR)~\cite{harris2019redcap,moore2019review,sirur2018we}. They provide structured data collection, data integrity, and privacy protection, allowing researchers to focus on generating meaningful insights without administrative or technical obstacles. The widespread use of RMAs highlights the need for robust security measures to maintain regulatory compliance and trust. Fig.~\ref{fig:rma} shows the basic workflow of RMAs.

Multi-center studies are increasing, growing complexities of data types have amplified the need for RMA integration. Platforms like REDCap, Castor, and Medidata Rave~\cite{van202435} have emerged as prominent solutions, varying in task performance but sharing similar architecture. Here, we focus on REDCap, it has become a go-to platform for secure data collection, supporting a variety of study designs, and enabling seamless integration with other research software tools~\cite{harris2009research,harris2019redcap}. It facilitates secure, web-based data collection and management. It has built-in features like role based access, encryption, and regular system updates to protect against unauthorized access, data breaches, and other security threats. However, REDCap needs to keep up with the emerging security threats in a landscape marked by evolving cyber-attacks and increasingly sophisticated malicious actors. Institutions need to implement added security to boost built in security to be protected against malicious users targeting the institution separately, like the AIDS Alabama data breach from a ransomware attack that occurred between October 11, 2021 and August 9, 2022~\cite{steven2023aids}.\\
For our findings broader context, we compare REDCap with RMAs like Castor and Medidata Rave. Castor offers similar features but emphasizes ease of use and rapid deployment. It is suitable for smaller projects. Medidata Rave~\cite{nizamicomprehensive} is designed for large-scale clinical trials and offers advanced data analytics and integration capabilities. We highlight the security challenges and mitigation strategies employed by RMAs.

\subsection{Threat Model}
A structured framework for identifying, assessing, and prioritizing potential security threats to a system, enabling the development of mitigation strategies~\cite{mahak2021threat}. It involves understanding the system's architecture, identifying potential threats, and determining appropriate countermeasures to reduce risks. It first identifies and categorizes key assets then evaluates the attack entry points, such as login pages, APIs, and network interfaces. Then defines the attack model by identifying potential threat actors like external hackers or malicious insiders. Specific threats and vulnerabilities are mapped such as SQL injection, weak password policies, and improper data access controls, which may lead to data confidentiality, integrity, or availability compromise. Finally, it identifies mitigation strategies to address possible threats such as implementing multi-factor authentication, regular software updates, and continuous monitoring. Table ~\ref{tab:keyaspects} shows the key aspects of threat modeling.

\begin{table}[!t]
\caption{Key Aspects of Threat Modeling}
\begin{center}
    \begin{tabularx}{\linewidth}{|p{1.35cm}|X|p{3cm}|}
        \hline
        \textbf{Aspect} & \textbf{Description} & \textbf{REDCap Example} \\
         \hline
         System Understanding & Understanding system's architecture and DFDs. & Mapping DFDs and identifying critical components.\\
         \hline
         Threat Identification & MITRE ATT\&CK methodology and STRIDE framework threat identification & Spoofing (unauthorized access), Tampering (data manipulation), DoS attacks.\\
         \hline
         Risk Assessment & Assessing threats based on impact and likelihood to prioritize mitigation efforts. & Evaluating the risk of data breaches and prioritizing high-impact threats. \\
         \hline
         Mitigation Strategies & Developing countermeasures to mitigate identified threats. & Enhance authentication, encrypt data, employ access controls, patch vulnerabilities. \\
         \hline
         Continuous Monitoring & Regularly reviewing and updating the threat model to adapt to new threats. & Ongoing security assessments and updating threat models. \\
         \hline
         Benefits & Enhancing security, ensuring regulatory compliance, and improving trust in the research environment. & Protecting sensitive data, meeting HIPAA/GDPR requirements, fostering a trustworthy environment. \\
         \hline
    \end{tabularx}
    \label{tab:keyaspects}
\end{center}
\vspace{-5pt}
\end{table}

\subsection{Current Regulatory and Data Trends}

The legal and ethical requirements governing RMAs are continuously evolving, moving toward stricter, patient-centric controls:

\begin{itemize}
    \item \textbf{GDPR and HIPAA Evolution:} Both HIPAA and GDPR continue to evolve, with mandates focusing on data minimization, where data collection is limited to only what is strictly essential for research. Under GDPR, consent must be explicitly given, and the right to be forgotten (data erasure upon request) is a key challenge for long-term clinical trial data retention~\cite{jurczuk2024consent}. 
    \item \textbf{De-identification and Pseudonymization:} Due to the risk of re-identification through data linkage, RMAs must rigorously employ pseudonymization (replacing identifiers with artificial ones) and advanced anonymization techniques, even for data shared outside the primary research team~\cite{garfinkel2015identification}.
    \item \textbf{Global Collaboration and Traceabilitiy:} The need for harmonized guidelines across countries is increasing due to multi-site studies. The FDA, for example, has moved to electronic submission systems to improve the traceability and accountability of every clinical trial data point.
\end{itemize}

\section{Assets}
\label{sec:assets}
RMAs are comprised of various assets critical to functionality, security, and usability. Identifying these assets helps us to understand potential threats and how to mitigate them:

\noindent \textbf{Research Data}: Sensitive data like personal health information (PHI) is a lucrative asset and must be protect to ensure privacy and security.\\
\noindent \textbf{Data Collection Instruments}: Researchers can create and customize data collection instruments like surveys and forms. Their flexibility tailors the data collection to specific research needs like various question types, branching logic, and data validation rules to ensure data accuracy and completeness~\cite{kianersi2021use}. \\
\noindent \textbf{Automated Workflows and Alerts}: These streamline the data collection process and improve efficiency by sending automated reminders to participants for surveys or notifying researchers of important events, ensuring timely data collection and participant engagement~\cite{kianersi2021use}.\\
\noindent \textbf{Data Access and Management}: Real-time data access allows researchers to monitor study progress and make timely decisions, especially for large-scale studies where immediate data access is critical for management and analysis~\cite{carmezim2024redcapdm}.\\
\noindent \textbf{Security and Compliance}: Data is protected and secure through features like encryption, role based access, and authentication. Also, REDCap is designed to comply with regulatory standards such as HIPAA, GDPR, and FDA~\cite{patridge2018research}.\\
\noindent \textbf{Multi-site and Multilingual Support}: RMAs support multi-site studies and multilingual capabilities suitable for international research projects, allowing collaboration across different locations, collecting data from diverse populations~\cite{mattingly2019towards}.\\
\noindent \textbf{Data Export and Integration}: With seamless data export to statistical software packages like SAS, SPSS, and R~\cite{patridge2018research}, RMAs integrate with systems like electronic health records (EHRs), enhancing their utility in clinical research~\cite{hawley2021digitization}.\\
\noindent \textbf{Mobile Access Point}: RMAs offer a mobile app, collects data remotely useful for field research or in areas with limited internet connectivity, ensuring uninterrupted data collection~\cite{chen2019evaluation}.

\section{Entry Points}
\label{sec:entry}
Entry points in RMAs are crucial for ensuring that access is secure and efficient. Methods and interfaces through which users can access and interact with the platform are many. These can also provide ways in which a malicious user may gain access and exploit system. Key RMA entry points are:

\noindent \textbf{User Authentication and Access Control}: RMAs ensure authorized access through robust user authentication strategies such as institutional credentials. External collaborators can be granted access via a sponsor. At the project level, the admin can configure access control, project owners can assign specific privileges to study staff according to their roles~\cite{patridge2018research}.\\
\noindent \textbf{Web Interface}: RMAs primary entry point is the secure web interface. It can be accessed from any connected device like laptops, smartphones, and tablets~\cite{chen2019evaluation}. This flexibility allows researchers to manage projects and enter data from virtually anywhere, enhancing the system accessibility and usability.\\
\noindent \textbf{API (Application Programming Interface)}: APIs allow for programmatic access to RMAs making them useful for integrating with systems like EHRs or external databases~\cite{kianersi2021use}. APIs support various functions, including data import/export, project management, and user management, enabling seamless interoperability with other research tools~\cite{kianersi2021use, patridge2018research}.\\
\noindent \textbf{Mobile App}: This serves as a remote data collection entry point. It can be set up to collect data whilst offline making it valuable for field research especially with limited internet connectivity. Data collected offline can be synchronized with the main server once an internet connection is available, ensuring continuous uninterrupted data collection~\cite{harris2021redcap}.\\
\noindent \textbf{Survey Links and Email Invitations}: REDCap allows researchers to distribute surveys via email invitations or public survey links for participant surveys~\cite{chen2019evaluation}. Participants can access these surveys through a secure link, providing a straightforward entry point for data collection. This enhances reaching a broad audience and facilitates participant engagement.\\
\noindent \textbf{Training and Support Resources}: Built-in training and support resources like tutorials, help text, and a comprehensive FAQ section can serve as entry points for gaining knowledge in understanding the system features and troubleshoot issues.

\section{Attack Model}
\label{sec:attack}
Attack model, a structured framework for defining traits, motivations, capabilities, and goals of potential threat actors who may target a system, application, or organization. Developing an attack model is essential for understanding the security risks associated with critical assets. Understanding the attackers and motivations is crucial for developing effective security measures in RMAs. By defining potential threats and their drivers, institutions can ensure the integrity of research activities through better data protection. We present a hypothetical example to illustrate the practical application of identified threats. In this scenario, an attacker attempts to gain unauthorized access through a phishing email to exploit a vulnerability and escalate privileges to exfiltrate data. This can be one of the many vulnerabilities faced by RMAs.

\subsection{Attackers}
\noindent \textbf{Cyber-criminals}: They often aim to exfiltrate data like PII or intellectual property, which is lucrative on the dark web. They may also deploy ransomware to extort money by encrypting critical research data and release it only if a ransom is paid~\cite{strom2018mitre}.\\
\noindent \textbf{Hacktivists}: They are driven by ideological or political beliefs. They may target RMAs to make a statement or protest against certain research activities or institutions. For example, they might deface websites or leak sensitive data to draw attention or damaging reputations more than financial gains~\cite{strom2018mitre}.\\
\noindent \textbf{Nation-State Actors}: Driven by espionage and strategic advantage, they target RMAs for data exfiltration, mainly in fields like health, biotechnology, pharmaceuticals, and defense. They are typically well-resourced and highly skilled, making their attacks sophisticated and difficult to detect~\cite{moller2023cyberattacker}.\\
\noindent \textbf{Insiders}: Disgruntled of former employees, may attack out of revenge or personal gain. Their legitimate access to the system can be used to exfiltrate or sabotage data. These are challenging to mitigate due to the attackers’ familiarity with the system and its security measures.\\
\noindent \textbf{Script Kiddies}: They use pre-written scripts or tools to exploit known vulnerabilities. While their attacks are often less sophisticated, they can still cause significant disruption.

\subsection{Attacker Capabilities}
Modern attackers, sometimes referred to as cyber-criminals and notion-state actors, are leveraging Artificial Intelligence (AI) to enhance their capabilities:\\
\noindent \textbf{Automated and Personalized Phishing}: Attackers are using generative AI to produce highly personalized and convincing phishing emails and deepfake voice/video calls that impersonate executives or clinicians, making them significantly harder to detect.\\
\noindent \textbf{AI-Driven Malware}: Malicious actors use AI tools to quickly write an compile new malware, accelerating the speed and complexity of the attack execution phase.\\
\noindent \textbf{Adversarial AI}: In research were RMAs utilize Machine Learning (ML) for advanced analytics, attackers can employ poisoning attacks to inject misleading data into the model's training dataset or evasion attacks to manipulate data input, thereby compromising the model's integrity and objectivity.\\
\noindent \textbf{Cyber-criminals}: Can steal data, plant ransomware, and use phishing attacks to gain unauthorized access to RMAs~\cite{lallie2021cyber}.\\
\noindent \textbf{Hacktivists}: Can disrupt operations, leak sensitive data, and deface RMAs to draw attention to their cause.~\cite{romagna2024becoming}\\
\noindent \textbf{Nation-State Actors}: Are capable of espionage, advanced persistent threat (APT) to continuously gain access undetected over a period of time, and sabotage research work~\cite{lallie2021cyber, jbair2022threat}.\\
\noindent \textbf{Insiders}: Are capable of data exfiltration, system manipulation, and credential sharing with external users~\cite{sharma2020user}.\\
\noindent \textbf{Script Kiddies}: Deploy basic exploits and simple attacks such as social engineering to deface or disrupt RMAs~\cite{burruss2022website}.

\subsection{Motivation}
\noindent \textbf{Financial Gain}: Common motivator for cyber-criminals targeting RMAs~\cite{strom2018mitre}. They seek to profit despite the harm they may cause like deploying ransomware or engaging in fraud.\\
\noindent \textbf{Ideological Beliefs}: Attackers such as hacktivists, aim to promote their political or social agendas. They may target research institutions conducting controversial studies or those affiliated with certain governments or corporations. Their goal is often to cause reputational damage or disrupt operations~\cite{strom2018mitre}.\\
\noindent \textbf{Espionage}: Attackers like nation-state actors, seek to gain strategic advantages by data exfiltration to advance their own research, gain economic benefits, or undermine competitors. They are typically sophisticated and well-funded~\cite{moller2023cyberattacker}.\\
\noindent \textbf{Revenge}: Attackers with system legitimate access like disgruntled insiders can be dangerous. They aim to harm the organization they feel wronged them by data exfiltration or destruction, or other forms of sabotage.\\
\noindent \textbf{Thrill and Notoriety}: Script kiddies are motivated by the thrill of hacking and the desire for recognition within their community targeting systems opportunistically, using readily available tools and scripts. While their attacks may lack sophistication, they can still cause significant disruption.

\section{Vulnerabilities}
\label{sec:vulnerabilities}
RMAs are critical for handling sensitive data that can have catastrophic repercussions if exposed. This makes them susceptible to various vulnerability attacks. Now we explore these vulnerabilities using Data Flow Diagrams (DFDs), the MITRE ATT\&CK framework, and the STRIDE model. 

\subsection{Data Flow Diagram (DFD)}
\begin{figure}[!t]
    \centerline{\includegraphics[width=2.8in,alt={REDCap Data Flow Diagram}]{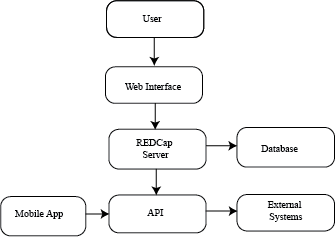}}
    \caption{REDCap Data Flow Diagram}
    \label{fig:dfd}
\end{figure}

DFDs are essential for visualizing the flow of data within a system focusing on critical components, and where data is processed, stored, or transmitted. Fig.~\ref{fig:dfd} shows the DFD for the REDCap RMA. Data is securely stored in secure database servers which can be vulnerable to SQL injection attacks if proper input validation is not implemented. Data is transmitted over networks between the users, REDCap server, and the database server. Unencrypted transmission can be intercepted causing data breaches. Data processing points can be targeted by attackers with buffer overflows or code injection attacks. 

\subsection{MITRE ATT\&CK}
The framework offers a panoramic matrix of strategies used by attackers to exploit vulnerabilities~\cite{strom2018mitre}. It helps understand how vulnerabilities can be exploited and what defensive measures to implement. Table~\ref{tab:threatmodel} shows potential vulnerabilities to RMAs using the MITRE ATT\&CK framework.

\begin{table*}[!t]
\centering
\caption{Threats and Vulnerabilities using the MITRE ATT\&CK framework}
\begin{tabularx}{\linewidth}{|p{2.3cm}|p{4.2cm}|X|}
    \hline
    \textbf{Attack} & \textbf{Attack Technique} & \textbf{Description} \\
     \hline
     Initial Access & Phishing & Using deceptive emails to trick users into revealing credentials or downloading malware. \\
     & Exploit Public-Facing Application & Exploiting web applications vulnerabilities for unauthorized access.\\
     \hline
     Execution & Command and Scripting Interpreter & Using command-line interfaces or scripts to execute malicious code.\\
     & PowerShell & Leveraging PowerShell to execute commands and scripts. \\
     \hline
    Privilege Escalation & Exploitation for Privilege Escalation & Exploiting vulnerabilities to gain higher-level permissions.\\
    & Bypass User Account Control & Circumventing User Account Control (UAC) to execute tasks with elevated privileges.\\
    \hline
    Defense Evasion & Obfuscated Files or Information & Hiding malicious code within obsfuscated files to avoid detection. \\
    & Disable Security Tools & Disabling antivirus or other security tools to evade detection. \\
    \hline 
    Credential Access & Credential Dumping & Extracting credentials from operating systems or applications. \\
    & Brute Force & Attempting to guess passwords through repeated trial and error. \\
    \hline
    Discovery & System Information Discovery & Gathering information about system, such as OS version and hardware details. \\
    & Network Service Scanning & Scanning for open network services to identify potential targets.\\
    \hline 
    Lateral Movement & Remote Services & Using remote services like RDP or SSH to move laterally within the network.\\
    & Pass the Hash & Using hashed credentials to authenticate and move laterally. \\
    \hline
    Collection & Data from Local Systems & Collecting data stored on the local system. \\
    & Input Capture & Capturing user input, such as keystrokes or clipboard data. \\
    \hline
    Exfiltration & Exfiltration Over C2 Channel & Sending data to an external server tasked with command and control (C2). \\
    & Automated Exfiltration & Data exfiltration using automated tools. \\
    \hline
    Impact & Data Destruction & Deleting or corrupting data to disrupt operations. \\
    & Data Encrypted for Impact & Encrypting data to render it unusable, often as part of a ransomware attack. \\
    \hline
\end{tabularx}
\label{tab:threatmodel}
\end{table*}

\subsection{STRIDE METHODOLOGY}
The methodology categorizes vulnerabilities into Spoofing, Tampering, Repudiation, Information Disclosure, Denial of Service, and Elevation of Privilege~\cite{khan2017stride}. It helps in identifying and addressing vulnerabilities systematically against which of the system components it would be vulnerable for attack. Table~\ref{tab:stride} highlights the STRIDE methodology for RMAs.

\begin{table*}[!t]
\begin{center}
    \caption{STRIDE Methodology for RMAs}
    \begin{tabularx}{\linewidth}{|p{3cm}|p{2cm}|p{4.5cm}|X|}
        \hline
        \textbf{Threat Category} & \textbf{Violates} & \textbf{Definition} & \textbf{Example} \\
         \hline
         Spoofing & Authenticity & Impersonating legitimate users or system to gain unauthorized access. & An attacker steals credentials or the authentication token of a legitimate user and uses it to impersonate the user.\\
         \hline
         Tampering & Integrity & Unauthorized modification of data or system configurations. & An attacker abuses the application to perform unintended updates to a database to manipulate research results.\\
         \hline
         Repudiation & Non-Repudiation & Denying the performance of an action without a way to prove otherwise.  & An attacker manipulates logs to cover their actions and denies submitting or altering data.\\
         \hline
         Information Disclosure & Confidentiality & Sensitive information unauthorized access. & Sensitive data is exposed due to improper access controls.\\
         \hline
         Denial of Service (DOS) Attacks & Availability & Disrupting service availability to legitimate users. & An attacker floods the server with requests, making the RMA unavailable.\\
         \hline
         Elevation of Privileges & Authorization & Gaining higher-level permissions than authorized. & An attacker exploits a vulnerability to gain admin access to the RMA.\\
         \hline
    \end{tabularx}

    \label{tab:stride}
\end{center}
\end{table*}

\subsection{Formal Threat Model: The Dolev-Yao Model}
The Dolev-Yao (DY) models is a formal model used to prove the security properties (secrecy, authentication, integrity) of interactive cryptographic protocols (e.g. TLS handshakes, API key exchanges) used in RMAs~\cite{herzog2005computational}. The DY model assumes the adversary has complete control over the network and can perform the following actions, limited only by the assumption that strong cryptographic primitives cannot be broken:
\begin{enumerate}
    \item \textbf{Intercept (Read):} Read and copy any message sent.
    \item \textbf{Fabricate (Inject):} Synthesize and inject new, forged messages.
    \item \textbf{Delete/Replay:} Prevent messages from reaching their destination or re-send old, legitimate messages.
    \item \textbf{Decrypt with Known Keys:} Decrypt any message if they possess the necessary key.
\end{enumerate}

\textbf{Dolev-Yao Application to RMA Data Flow}\\
This model is critical for analyzing the non-repudiation and confidentiality of data at the protocol level, especially between the user, web interface, RMA (REDCap) server, and database data flows as shown in Table~\ref{tab:dy}.

\begin{table}[!t]
\begin{center}
    \caption{Dolev-Yao Threat Model}
    \begin{tabularx}{\linewidth}{|p{2cm}|p{2cm}|X|}
        \hline
         \textbf{Component/Data Flow} & \textbf{STRIDE Property Violated} & \textbf{Dolev-Yao Attack Scenario}  \\
         \hline
         API/WEB Interface Key Exchange & Confidentiality (Information Disclosure & If the session key exchange protocol is flawed, the DY attacker can intercept and successfully derive the key, thereby compromising the secrecy of the sensitive data in transit\\ 
         \hline
         Data Transmission (API Integration) & Integrity (Tampering) & The DY attacker modifies or fabricates a message that simulates a successful data ingestion from an external EHR system to the REDCap database, leading to manipulated research results.\\
         \hline
         Automated Workflows (Survey Links) & Authentication (Spoofing) & The DY attacker replays a legitimate, time-sensitive survey link intended for a participant to gain unauthorized access to the study, leading to information disclosure or data tampering.\\
         \hline
    \end{tabularx}
    \label{tab:dy}
\end{center}
\end{table}

\section{Mitigation Strategies}
\label{sec:mitigation}

\subsection{Security Controls}
These help to fortify system defenses by safeguarding RMAs from security risks ensuring research data integrity, availability, and confidentiality. We now turn our attention to mitigation strategies, case studies and simulations for RMAs ensuring compliance with HIPAA~\cite{moore2019review} and effectiveness.\\
\noindent \textbf{Initial Access:} Employ solutions to filter emails to block and report phishing emails. Train users to recognize phishing attempts and report suspicious emails. Troubleshoot vulnerabilities by performing regular updates to patch the application. Implement firewalls to defend against common web exploits. This has shown to greatly reduce phishing incidents.\\
\noindent \textbf{Execution:} Restrict the use of scripting languages and command-line interpreters to authorized users only. Implement application whitelisting to prevent unauthorized script execution and maintain system integrity.\\
\noindent \textbf{Privilege Escalation:} Use a kernel observer to monitor during system call processing for any changes in privileges. This detects abnormal privilege changes and deploys countermeasures to prevent this in case of an attack~\cite{yamauchi2021additional}. Deployment of a kernel observer can successfully detect and mitigate privilege escalation attempts, ensuring secure operations.\\
\noindent \textbf{Defense Evasion:} Implement techniques like continuous monitoring and machine learning based techniques to detect malware and hidden code~\cite{hossain2024enhanced}. ML techniques can detect previously undetected malware in RMAs with accuracy.\\
\noindent \textbf{Credential Access:} Multi-factor authentication (MFA) can reduce the risk of credential theft, it is proven in the real world that is hard to steal all modes of authentication simultaneously.\\
\noindent \textbf{Discovery:} Grant privileges on a need to know basis using the principle of least privileges~\cite{kwon2020cyber}, unauthorized access attempts.\\
\noindent \textbf{Lateral Movement:} Segment the network to stop attackers from moving laterally. This significantly reduces lateral movement attempts, it preserves the integrity of the entire network.\\
\noindent \textbf{Collection:} Only authorized channels are supposed to connect to the database and perform transactions, this prevents unauthorized data collection attempts as proven in RMAs.\\
\noindent \textbf{Exfiltration:} Employ machine ML exfiltration techniques to aid exfiltration detection~\cite{sabir2021machine}. They have high success rates in identifying and preventing data exfiltration attempts.\\
\noindent \textbf{Impact:} Regularly backup data and ensure it is stored offline or on a separate network. Use endpoint detection and response (EDR) solutions to detect and block ransomware activities. This has greatly reduced the impact of ransomware attacks on RMAs because the organization still has backups of the data seized by attackers, operations can continue uninterrupted.\\
\noindent \textbf{Spoofing:} Implement strong authentication mechanisms like MFA and digital certificates, to verify user identifications. \\
\noindent \textbf{Tampering:} Use access control and cryptographic techniques like digital signatures or hash functions for data integrity. Tampered data will show digital signature modifications.\\
\noindent \textbf{Repudiation:} Logging and auditing mechanisms to track user actions and ensure user action accountability and tractability.\\
\noindent \textbf{Information Disclosure:} Encrypt sensitive data and only allow data access to authorized users.\\
\noindent \textbf{Denial of Service (DOS) Attacks:} Rate limiting and network traffic filtering prevent DoS attacks. Use redundant systems and load balancing to ensure service availability. Redundant systems act like data backups ensuring organization seamlessly switches to the other system in case of an attack.\\
\noindent \textbf{Elevation of Privileges:} Regular system updates to fix vulnerabilities preventing privilege escalation and boost security.\\
Based on our example in the Section V, the proposed strategies prove to neutralize identified threats. Such as, email filtering and user training can prevent phishing attempts. System updates and firewalls can defend against web exploits. MFA and whitelisting can restrict unauthorized access and script execution. 

\subsection{Advanced Identity and Architectural Controls}
Mitigation strategies must move beyond perimeter defenses to address identity-centric and protocol-level threats:\\
\noindent\textbf{Mandatory Multi-Factor Authentication (MFA):} The incident at the \textbf{Change Healthcare} proved that a single point of failure, a Citrix portal lacking MFA, can lead to the compromise of data for over 190 million individuals. MFA must be enforced on all access points, including all remote access applications, third-party vendor connections, and privileged accounts.\\
\noindent\textbf{Zero Trust Architecture (ZTA):} RMAs should adopt ZTA, operating on the principle of "\textit{never trust, always verify}". This means:\\

\begin{itemize}
    \item \textit{Continuous Monitoring:} Every access request, whether from inside or outside the network, must be continuously authenticated an authorized based on context (user, device posture, location).
    \item \textit{Microsegmentation:} Networks must be segmented to isolate critical systems (e.g., separating the database server from the administrative workstations). This limits the ability of an attacker to move laterally even if they breach an initial segment.
    \item \textit{Least Privilege:} Users and devices are granted the absolute minimum level of access required for their role, which limits the "blast radius" of compromised credentials.
\end{itemize}

\subsection{Formal Protocol Verification}
To counter the Dolev-Yao model's powerful adversary, communication protocols, especially those involving API data transfer and secure session establishment, should be formally modeled using tools like ProVerif. This tool uses the Dolev-Yao model to automatically check for vulnerabilities like key leaks, replay attacks, and authentication flaws, proving the protocol's security properties before deployment~\cite{blanchet2018proverif}.

\section{Related Works}
\label{sec:work}
RMAs are a critical area of study, in clinical research, REDCap is popular for having robust security features and HIPAA and GDPR compliance~\cite{harris2019redcap,harris2021redcap,moore2019review,sirur2018we}. Odukoya et. al affirms the importance of secure data management tools in research activities like clinical trials~\cite{odukoya2021application}. Security management tools can be expensive for smaller projects, RMAs with low costs and minimal programming requirements can be options for such research projects~\cite{parkinson2022survey}. Tariq e. al. discusses cybersecurity challenges and future research directions for IoT applications which share similar concerns with RMAs~\cite{tariq2023critical}.

\section{Conclusion}
\label{sec:conclusion}
In this research, we analyzed the vulnerabilities and security risks faced by Research Management Applications (RMAs) focusing on REDCap, using Data Flow Diagrams (DFD), MITRE ATT\&CK framework, and STRIDE methodology. We analyzed various aspects of RMA security, identified assets, critical components, and areas of vulnerability. We defined the possible threats such as cyberattacks, insider attacks, and data collection and transmission vulnerabilities which can compromise the system resulting in data breaches, and described mitigation strategies for reducing risk. For clinical research institutions, RMA protection is paramount. Our study showed various mitigation strategies such as email filtering solutions, authorized user access, system monitoring, automatic malware detection, MFA, and cryptographic techniques. Institutions can vastly improve security by combining different security techniques at data collection, transmission, and storage levels. We plan to leverage machine and deep learning algorithms in enhancing our processes and solutions~\cite{sarker2021ai,sarker2023internet} in future.

\balance

\end{document}